# Autonomous Optimization of Complex Oxides for Thermochemical Fuel Production

Shuiping Gong[1,2], Mingcheng Li[3], Han Hao[4], Zhenhao Zhou[1,2], Yi Li[1,2], Xiaobo Liao[1,2], Cheng Fang[1,2], Jian Deng[1,2], Jiangang He[5,6], Wenpei Gao[1,2], Yakun Yuan[2], Tao Deng[1], Chaochao Dun[7], Runxia Cai[3], Zhenpeng Yao[1,2,4]

[1]Center of Hydrogen Science and School of Materials Science and Engineering, Shanghai Jiao Tong University, Shanghai 200240, China

[2]Innovation Center for Future Materials, Zhangjiang Institute for Advanced Study, Shanghai Jiao Tong University, Shanghai 201203, China

[3]School of Mechanical Engineering, Shanghai Jiao Tong University, Shanghai, China

[4]Acceleration Consortium, 700 University Avenue, Toronto, Ontario M7A 2S4, Canada

[5]Key Laboratory of Advanced Materials and Devices for Post-Moore Chips, Ministry of Education, University of Science and Technology Beijing, Beijing 100083, China

[6]School of Mathematics and Physics, University of Science and Technology Beijing, Beijing 100083, China

[7]Department of Chemical and Biological Engineering, University at Buffalo, The State University of New York, Buffalo, NY 14260, United States of America

## ABSTRACT

Two-step thermochemical cycles for $H_2O$ and $CO_2$ splitting offer a promising route to sustainable fuel production. Their performance depends on redox-active oxides that undergo cyclic reduction and oxidation reactions. Achieving high thermal-to-fuel conversion efficiency requires both favorable redox thermodynamics and sufficiently rapid reaction kinetics. These requirements have shifted materials design toward complex multication oxides, including mixed-cation fluorites, perovskites, and high-entropy oxides. Developing such materials requires the simultaneous optimization of their composition, defect chemistry, phase stability, and morphology. The resulting design space is difficult to explore through sequential trial and error. Thermochemical fuel production therefore provides a compelling testbed for autonomous materials design and optimization. In this study, we first examine why redox-active complex oxides are challenging to develop owing to their multidimensional phase spaces, harsh operating conditions, and competing functional targets. We then discuss how high-throughput computation, automated synthesis, characterization, and testing, together with machine learning, can be integrated into closed-loop workflows to address these challenges. Drawing on advances from broader oxide materials research, we define a capability roadmap encompassing complex-oxide synthesis, *operando* characterization, robotic testing, condition-aware computation, and multi-objective optimization. Finally, we identify the key experimental, computational, data, and artificial intelligence capabilities required to build self-improving optimization platforms for materials development in thermochemical fuel production.

## INTRODUCTION

Thermochemical fuel production has emerged as a promising pathway for the cyclic conversion of high-temperature heat, such as concentrated solar[1–4], nuclear reactor[5–8], or high-grade industrial waste heat [9,10], into chemical fuels such as hydrogen and syngas. Among the proposed thermochemical cycles[11–13], two-step cycles offer high theoretical energy efficiencies[14–16], separate fuel and oxygen generation processes[17–19], and a reduced reliance on precious elements[20,21]. In a typical two-step cycle, a metal oxide undergoes endothermic thermal reduction at high temperatures (typically ~1000–2000 K, depending on the oxide chemistry) to release oxygen, followed by lower-temperature exothermic re-oxidation with $H_2O$ or $CO_2$ to produce $H_2$ or CO [22–24] while regenerating the oxide.

A wide range of metal oxides has been investigated for two-step thermochemical fuel production. Based on their redox mechanisms and structural evolution, these oxides are commonly categorized as stoichiometric systems (*e.g.*, ZnO/Zn)[25,26], and nonstoichiometric systems (*e.g.*, $CeO_{2-\delta}$ and $CaTi_{0.5}Mn_{0.5}O_{3-\delta}$)[23,27]. Despite substantial efforts to improve redox-active metal oxides, achieving commercially competitive performance remains challenging. For example, the U.S. Department of Energy has established ultimate targets of 26% solar-to-hydrogen efficiency and a hydrogen production cost of 2 USD/kg for solar thermochemical $H_2$ production[28]. Reactor architectures and heat recovery also strongly influence overall system performance and have been reviewed elsewhere[29–31]. Here, we focus on the materials constraints governing redox thermodynamics, reaction kinetics, and cycling stability, which collectively limit progress toward the performance targets required for practical deployment.

At the materials level, maximizing the two-step cycle performance requires redox-active materials that simultaneously satisfy multiple criteria, including low reduction temperatures[32,33], high specific fuel yields[34,35], rapid reaction rates[36,37], resistance to sintering-induced deactivation[38,39], and suppressed thermal volatilization[34,40,41]. Compositional engineering has been widely employed in both stoichiometric and nonstoichiometric metal oxides to balance these properties[24,42–45]. This approach has increased interest in multication oxides, including mixed-cation fluorites, perovskites, and high-entropy oxides[11,46–48]. Nevertheless, composition alone is insufficient to achieve optimal performance. The defect chemistry[49,50], phase thermodynamics[34,49,51], and morphology[52,53] also strongly influence the reaction thermodynamics, kinetics, and cycling stability.

These coupled variables make the optimization of redox-active oxides particularly challenging. First, complex oxide-based redox-active materials occupy a large compositional and structural space[54,55]. For example, double perovskites ($AA'BB'O_6$), one of the most extensively studied redox-active oxide families[44,56], permit millions of nominal cation combinations[57]. Defects, phase assemblages, and morphology add further variables. Second, thermochemical fuel production imposes severe operating conditions across distinct temperature windows. Temperatures above 1000 K are typically indispensable to obtain a sufficient reduction driving force and oxygen-exchange rates[13,14,58]. Under these conditions, however, redox-active oxides are susceptible to sintering, phase transformations, volatilization, and mechanical degradation during repeated redox cycling[59,60]. These extreme operating conditions also complicate testing and *operando* characterization, substantially slowing materials development. Third, redox-active oxides exhibit inherent property trade-offs[40,61]. For example, weaker metal–oxygen bonding can lower the reduction temperature[62–64] but may also reduce the driving force for re-oxidation. A greater oxygen capacity can increase the fuel yield but may compromise the structural stability[65,66]. These constraints limit the effectiveness of sequential, intuition-driven screening, which samples only a small fraction of the available design space[13,67,68]. Conventional trial-and-error approaches are therefore poorly matched to the scale and complexity of the redox-active oxide design problem because they are sequential, low-throughput, and often biased toward familiar compositional families.

Data-driven materials discovery, including self-driving laboratories (SDLs)[69,70], artificial intelligence (AI)[71,72], and high-throughput computational materials science[73–75], offers an opportunity to expand the range of compositions and operating conditions that can be explored[57,76,77]. SDLs combine automated hardware with decision-making algorithms to execute iterative cycles of synthesis, characterization, and testing[70,78]. Machine-learning (ML) models can rank candidates based on measured or computed descriptors and enable inverse design toward specified target properties[71,79]. High-throughput calculations can screen the phase stability, defect energetics, and kinetic barriers relevant to redox cycling, thereby narrowing the experimental focus to the most promising regions of the chemical space[50,80]. When integrated, these approaches can transform redox-active materials discovery from a slow, trial-and-error process into a predictive, continuously learning, closed-loop workflow capable of navigating complex chemical spaces under harsh operating conditions.

In this materials-centered review, we move beyond a catalog of automation technologies applied to thermochemical fuel production and instead organize the field around the core paradigms required for accelerated redox-active materials design and optimization. We first examine the fundamental materials bottlenecks that have hindered progress, focusing on multidimensional phase spaces, high-temperature degradation, and intrinsically multi-objective performance trade-offs. We then critically assess advances in automated and high-throughput synthesis, characterization, computation, and ML, emphasizing how these approaches enable the rapid exploration of design spaces, learning from realistic cycling data, and the co-optimization of the composition, processing, and performance. Finally, we outline the primary experimental, computational, data, and AI challenges that must be overcome to build closed-loop, self-improving discovery platforms. Thermochemical fuel production provides a stringent testbed for autonomous methods because its experiments are high-temperature, cyclic, and history-dependent.

## DESIGN CHALLENGES OF REDOX-ACTIVE COMPLEX OXIDES

Redox-active oxide development involves more than selecting an optimal composition. Candidate materials must satisfy multiple coupled performance criteria while retaining their functionality under repeated thermal and atmospheric cycling (**Fig. 1**).

*Enormous multidimensional phase spaces.* Multication oxides permit substitution across multiple crystallographic sites and therefore span many nominal compositions. At the compositional level, candidate materials such as doped fluorites, double perovskites, mixed-cation spinels, and emerging high-entropy oxides allow extensive cation substitution across multiple crystallographic sites, generating a large number of possible compositions and configurations[81–83]. Even at a fixed nominal composition, oxygen nonstoichiometry[45,84], vacancy ordering[85,86], cation ordering[87,88], and secondary-phase formation can vary with the synthesis history[89,90], and operating conditions[13,51]. Processing conditions further influence the particle size, porosity, surface area, grain structure, and phase assemblage[90,91], all of which can affect the reaction thermodynamics, kinetics, and degradation during high-temperature cycling[34,53,92]. As a result, complex oxides for two-step thermochemical fuel production occupy a multidimensional composition–defect–structure–microstructure space that is far too large to explore efficiently using conventional trial-and-error approaches. Generative AI could further expand this search space by proposing candidate compositions or structures subject to specified constraints. These candidates can then be screened using ML-based property predictors, density functional theory (DFT) calculations, and thermodynamic criteria before experimental validation, potentially accelerating the discovery and optimization of high-performance redox-active oxides[93–95].

*Harsh operating conditions*. Thermochemical fuel production inherently requires high-temperature operation, typically in the range of 1200–1700 K, to establish the necessary thermodynamic driving force and to facilitate oxygen-exchange kinetics. Consequently, redox-active oxides are subjected to severe thermal and mechanical stresses[96,97]. Repeated cycling can induce lattice strain, microcracking, fragmentation, and, ultimately, mechanical failure[96,98]. Moreover, at these temperatures, sintering driven by atomic-scale diffusion promotes grain coarsening and particle coalescence, which collectively reduce gas–solid contact and reactivity. In fluidized systems, severe agglomeration can also cause defluidization[99,100]. In multicomponent or composite oxide architectures, unequal cation diffusivities can promote segregation *via* Kirkendall-type processes and weaken designed composite structures during cycling[101,102]. Because these degradation mechanisms are strongly coupled, performance loss cannot be attributed to a single variable[101,103]. Moreover, high-temperature testing under controlled atmospheres is inherently slow and resource-intensive, particularly when long-term cycling and *operando* measurements are required.

*Competing functional targets*. Redox-active oxides must simultaneously satisfy multiple functional objectives, many of which are inherently coupled or competing. Several examples illustrate these trade-offs. a) Reduction temperature *versus* re-oxidation driving force: Lowering the reduction temperature generally requires more labile lattice oxygen or weaker effective metal–oxygen bonding. However, excessive reducibility can decrease the thermodynamic driving force for subsequent re-oxidation by $H_2O$ or $CO_2$, thereby limiting fuel production[62,104]. b) Oxygen capacity *versus* cyclability: In nonstoichiometric oxides, a high oxygen capacity is desirable because greater oxygen release can increase the specific $H_2$ or CO yield. Excessive nonstoichiometry, however, can destabilize the crystal framework, induce phase transitions, and compromise cyclability[39,105]. c) Kinetics *versus* sintering: Rapid fuel production requires fast surface exchange and oxygen transport. However, the high defect mobility and mass transport at elevated temperatures that facilitate rapid redox kinetics can also accelerate sintering, cation migration, and microstructural coarsening[91,106]. d) Surface reactivity *versus* particle passivation: Highly reactive surfaces can enhance $H_2O$ or $CO_2$ activation but may also be more susceptible to surface reconstruction, segregation, impurity poisoning, or irreversible passivation[65,107]. Optimizing redox-active oxides therefore requires careful tuning of the stoichiometry, composition, crystal structure, phase assemblage, and morphology to balance the thermodynamics, kinetics, and durability. These coupled trade-offs complicate property prediction and render conventional trial-and-error optimization slow, resource-intensive, and difficult to scale.

Together, these challenges explain why conventional trial-and-error approaches are poorly matched to the optimization of redox-active oxides. Sequential, low-throughput, and intuition-driven workflows sample only a small fraction of the available chemical space. Advancing thermochemical fuel production therefore requires a discovery framework that can explore broad compositional landscapes and test materials under relevant operating conditions. Autonomous, data-driven materials discovery offers such a framework, motivating the roadmap developed in the following sections.

## AUTONOMOUS, DATA-DRIVEN MATERIALS DISCOVERY

Autonomous, data-driven materials optimization links models, automated experiments, and decision rules in an iterative workflow. This approach is useful when the search space is large, and each experiment is costly. Thermochemical fuel production represents such a domain. The workflow uses computational and experimental results to update a model and select subsequent experiments. (**Fig. 1**)[69,70,108].

To distinguish automation from autonomy and assess the maturity of emerging discovery capabilities, we propose a five-level operational framework (levels L0–L4) for redox-active oxide discovery. The framework is presented as an analytical classification rather than an international standard. Similar distinctions between automated execution, model-guided experimentation, and autonomous closed-loop discovery have been discussed in the broader context of self-driving laboratories and autonomous experimentation[109]. L0 represents manual experimentation with human-driven design, synthesis, characterization, and evaluation. L1 corresponds to automated execution, in which individual experimental steps are accelerated through robotic synthesis, automated characterization, or programmable testing. Examples include the automated scanning transmission electron microscopy (STEM) workflow reported by Olszta et al.[110] and the machine-learning-assisted single-crystal neutron diffraction workflow reported by Hao et al.[111] . Although these approaches incorporate automated image or diffraction data acquisition and analysis, their automation is primarily confined to individual characterization workflows rather than the selection and optimization of materials or synthesis conditions. L2 introduces model-guided discovery, in which computational or ML models rank candidates and prioritize experimental exploration, while experimental decisions remain human-directed. Examples of this approach include high-throughput computational screening of oxide materials using DFT- and ML-based predictions of oxygen-related properties[75,108,112]. L3 represents closed-loop optimization, in which experimental outcomes are automatically integrated into predictive models that determine subsequent experiments through active learning or adaptive decision-making. Such closed-loop strategies have been demonstrated in autonomous chemistry and materials discovery platforms, where ML algorithms select informative experiments based on uncertainty estimation and experimental feedback[76,113]. L4 describes self-directed discovery, in which autonomous systems formulate objectives, navigate multiple design spaces, and identify new materials or mechanisms with minimal predefined human intervention. Although such capabilities have been demonstrated conceptually in broader materials-discovery environments, they remain beyond the current state of thermochemical redox materials optimization.

Based on currently available literature, thermochemical redox materials discovery remains predominantly at levels L1 and L2, characterized by automated synthesis, high-throughput characterization, and computation-guided candidate screening. Recent studies have demonstrated ML-assisted identification and experimental validation of thermochemical materials[49,114–116], but these workflows still rely heavily on human-guided selection and validation rather than fully autonomous optimization. Several L3-compatible methods, including autonomous experimentation, active-learning-guided optimization, and adaptive characterization, have been demonstrated mainly in adjacent materials systems[76,117]. However, their integration under high-temperature redox cycling conditions remains limited. In particular, recent advances in *operando* characterization have begun to reveal the dynamic structural evolution of redox-active oxides during thermochemical operation, but no end-to-end L4 autonomous platform has yet been established for high-temperature cyclic oxide optimization under realistic thermochemical fuel-production conditions.

Importantly, these operational levels do not imply that every material descriptor must be measured using *in-situ* or *operando* techniques under full reaction conditions. L2 model-guided discovery can be supported by computational predictions, standardized *ex-situ* characterization, and reliable cyclic performance measurements, provided that the data are sufficiently reproducible, comparable, and accompanied by appropriate uncertainty estimates. High-temperature *operando* characterization is therefore not a universal prerequisite for L2 candidate prioritization. For L2 screening tasks whose ranking criteria depend on cycle-resolved phase or defect evolution, reliable high-temperature *operando* data remain a limiting factor; the maturity of L2 is therefore descriptor-dependent rather than uniform across thermochemical systems. Its limitations become particularly consequential for L3 workflows when subsequent experimental decisions depend on resolving dynamic changes in phase assemblage, defect chemistry, or microstructure during redox cycling. For such applications, reliable closed-loop operation requires not only automated experiments, but also calibrated measurements and robust data-processing methods capable of returning physically meaningful descriptors under representative operating conditions.

Within such a framework, the *large design space* of redox-active oxides becomes more tractable. High-throughput computations can rapidly evaluate large families of candidate oxides using screening descriptors such as phase stability, redox energetics, oxygen-vacancy formation, oxygen transport, and finite-temperature thermodynamics. These calculations can reduce the number of candidates advanced to experimental validation. Automated platforms can prepare and evaluate shortlisted candidates under common protocols, improving comparability across experiments. AI and ML can integrate these computational and experimental data streams[118,119] and select subsequent experiments to maximize information gain or progress toward target properties, thereby accelerating discovery and improving resource utilization[120,121].

Autonomous, data-driven approaches can also help address the *harsh operating conditions* of thermochemical fuel production by increasing the throughput, reproducibility, and automation of testing and characterization. At present, evaluating redox-active oxides under relevant conditions remains slow and resource-intensive because experiments require high-temperature heating and cooling, controlled gas atmospheres, sealed or well-isolated reaction environments, repeated redox cycling, and frequent sample loading, unloading, and post-test analysis[32,65,122]. Miniaturized reactors or test capsules, parallel testing architectures, and robotic sample handling can increase the number of materials evaluated per unit time, reduce sample consumption, and make large-scale compositional screening more practical[70,123]. Automation also enables tighter control over temperature programs, gas switching, cycling protocols, and data acquisition, improving comparability across experiments[113,121]. More importantly, automated testing and characterization provide an operational foundation for closed-loop materials optimization, in which algorithms can use experimental feedback to adjust the compositions, processing conditions, and operating parameters to select subsequent experiments with minimal human intervention[76,124,125].

Autonomous optimization is also well suited to balancing the *competing functional targets* in redox-active oxides design. Rather than optimizing a single descriptor, AI-guided workflows can quantify the trade-offs between multiple design targets, such as the reduction temperature, oxygen storage capacity, re-oxidation and fuel production, transport kinetics, and stability[76,126]. Uncertainty-aware models and active-learning strategies can prioritize maximally informative experiments, while multi-objective optimization can guide exploration toward Pareto-optimal candidates that balance performance and robustness[124,127]. By explicitly learning these trade-offs, autonomous materials discovery enables redox-active oxides to be optimized as integrated, evolving material systems rather than as materials with static compositions selected by isolated descriptors[70,76,128].

Therefore, a self-driving laboratory transitions from an automated test stand to a truly autonomous system only when the experimental results update the core model and are actively used to select subsequent experiments (**Fig. 1**). For thermochemical fuel production, this shift provides a scalable route to accelerate redox-active complex oxide discovery while embedding chemical complexity, harsh operating conditions, and multi-objective performance targets directly into the design loop. The roadmap presented below is organized around the capabilities required to implement this loop for redox-active oxides.

## CAPABILITY ROADMAP FOR AUTONOMOUS COMPLEX OXIDE DISCOVERY

Although autonomous discovery remains at an early stage in thermochemical fuel production, many of the capabilities required for the development of redox-active oxides have already emerged in adjacent areas of oxide materials research. These advances provide a transferable foundation for accelerating the design and optimization of redox-active oxides. In this section, we organize recent progress by capability rather than by technique, emphasizing how approaches developed for broader oxide discovery can be adapted to address the specific demands of thermochemical fuel production.

### Accelerated synthesis of complex oxides

Autonomous optimization requires the reproducible preparation of oxide libraries with controlled compositions, phases, and morphologies, while maintaining consistent synthesis outcomes across batches. Conventional synthesis of redox-active oxides, particularly *via* solid-state reactions, remains slow because it often relies on repeated grinding, calcination, and manual handling. Recent advances in automated and high-throughput oxide synthesis offer routes to overcome this bottleneck, as demonstrated in studies of high-throughput solid-state workflows, self-driving laboratories, and ML-accelerated approaches[129–133]. Ultrafast sintering[129], automated slurry-based solid-state synthesis[130] (**Fig. 2a**), robotic solid-state workflow[131], programmable flash sintering[134], continuous-flow solvothermal synthesis[135], automated sol-gel platforms[132], automated flame spray pyrolysis[133,136], and high-throughput coprecipitation[137] have all demonstrated the capacity to accelerate oxide preparation across diverse chemistries. These methods allow for more rapid variations in composition and processing conditions than conventional batch workflows. Several techniques also permit synthesis outside conventional equilibrium processing windows[138–141]. However, these approaches differ substantially in throughput and reproducibility, ranging from second-scale ultrafast sintering to minute- or hour-scale automated synthesis workflows. Some approaches use automation and parallelization to substantially improve reproducibility and the overall throughput, although the actual rate remains strongly dependent on the precursor preparation, synthesis, and post-treatment steps. Therefore, the choice of an accelerated synthesis method depends on the materials chemistry and operating requirements of the specific application.

High-throughput automated syntheses have been established for many binary and conventional oxides *via* liquid- or vapor-phase methods; currently, such closed-loop robotic workflows are being extended to multicomponent complex oxides[133,142,143]. Some of these methods have already been applied to redox-active-oxide-relevant systems (**Table 1**)[144,145], whereas others have mainly been demonstrated for functional oxides beyond thermochemical fuel production. For redox-active oxide discovery, the key opportunity is to adapt these synthesis platforms to generate libraries that systematically vary the cation identity, dopant concentration, phase assemblage, particle morphology, and processing history. Such control is essential because performance depends not only on the nominal composition, but also on the defect population, phase purity, surface area, porosity, and microstructural evolution during redox cycling. Therefore, the composition and synthesis parameters should be treated as joint design variables in a closed-loop campaign.

## Automated characterization of redox-active oxides

Autonomous optimization also requires characterization that is reproducible and fast enough to keep pace with synthesis (**Fig. 2b**). Composition and stoichiometry determine the redox thermodynamics, oxygen capacity, and degradation pathways, making techniques such as inductively coupled plasma[146], X-ray fluorescence (XRF)[147], X-ray photoelectron spectroscopy (XPS)[148,149], energy-dispersive X-ray spectroscopy[92], wavelength-dispersive spectroscopy[150,151], and secondary ion mass spectrometry[152,153] central to interpreting performance. Structural characterization by X-ray diffraction[78,154], neutron diffraction[111,155], transmission electron microscopy (TEM)[156,157], scanning transmission electron microscopy (STEM) [158,159], and Raman spectroscopy[160,161] provides complementary information on the phase identity, lattice evolution, oxygen-vacancy ordering, local coordination, and defect chemistry. Valence-state probes, including X-ray absorption near-edge structure[27,162], XPS[163,164], titration (*e.g.*, iodine titration and Coulomb titration)[165–167], and electron energy loss spectroscopy[157,168], reveal the redox-active cations and electronic changes that drive oxygen exchange. Morphological tools, including scanning electron microscopy[169,170], TEM/STEM[171,172], and Brunauer–Emmett–Teller analysis[102], track particle coarsening, porosity, and surface area, and comparisons before and after cycling can reveal coarsening and microstructural damage. For autonomous use, the characterization quality must therefore be judged not only by the information content, but also by the repeatability, calibration, sample representativeness, analysis robustness, and availability of machine-readable outputs with traceable uncertainty.

The broader oxide community has made substantial progress in automating characterization workflows (**Table 1**), including integrated XRD/XRF platforms[147], automated spectral fitting[173], robotic sample handling[167], ML-assisted diffraction analysis[174], automated Raman classification[161], and AI-enabled microscopy[158]. However, readiness for autonomous integration varies markedly between these techniques. Routine *ex-situ* XRD and XRF, together with robotic sample handling and automated diffraction analysis, are relatively mature for library-scale quality control[147,158,160]. Automated Raman analysis and AI-assisted microscopy are emerging[158,160], whereas TEM and STEM remain limited for broad screening by the sample preparation, field-of-view selection, beam effects, and data-analysis requirements. Therefore, the current level of automation and throughput varies substantially across characterization techniques. The optimization of complex oxides does not require high-throughput screening with every characterization technique. A tiered characterization pipeline can therefore realize a balance between the experimental cost and efficiency in the optimization of complex oxides. However, translating these capabilities to thermochemical fuel production remains challenging because the most informative measurements often require high temperatures, controlled oxygen partial pressures, and reactive gas atmospheres[154,166]. The next step is therefore not only faster characterization, but also tiered characterization pipelines that combine rapid *ex-situ* screening with targeted *operando* and post-cycling analysis. In a rational oxide screening strategy, the first layer of experimental screening can employ rapid tools such as automatic *ex-situ* XRD and XRF to characterize broad libraries of freshly prepared samples, identify unintended phases or large compositional deviations, and generate baseline structural and compositional descriptors for computational and ML models. Subsequently, candidates can be further screened by lower-throughput characterization techniques, such as XPS, to obtain more detailed property information. Finally, resource-intensive *operando* and post-cycling measurements, such as *in-situ* STEM, can be reserved for selected candidates, where they provide deeper insight into structure–performance relationships and cycling-induced evolution. In closed-loop workflows, these measurements should be converted into machine-readable descriptors, such as the elemental ratios, phase fractions, lattice parameters, cation valence states, oxygen nonstoichiometry, surface carbonate or hydroxyl signatures where relevant, and particle-coarsening metrics[175–177]. For closed-loop use, automated analysis should return both a descriptor and its uncertainty; qualitative labels alone are insufficient[78,117,158,173]

These challenges are particularly pronounced for *in-situ* and *operando* measurements under cyclic, reactive conditions at temperatures above approximately 800 °C. Such measurements may require specialized sample environments, high-temperature-compatible windows or capillaries, controlled gas delivery, and accurate temperature calibration[178,179]. Thermal gradients, changes in sample emissivity, and temperature-dependent background signals can complicate quantitative data interpretation. In infrared and Raman spectroscopy, thermal emission and background fluctuations may obscure weak spectral features[180,181], whereas in extended X-ray absorption fine structure (EXAFS) measurements, increased thermal disorder can dampen the oscillation amplitudes and complicate local-structure analysis[182,183]. Contributions from the sample environment may further affect signal quality. Although specialized setups can enable measurements under demanding conditions, their feasibility and data quality remain dependent on the technique, sample environment, and operating conditions.

These limitations have direct implications for autonomous experimentation because closed-loop decisions depend on the reliability and interpretability of the data returned by the measurement system. *Operando* characterization should therefore be selected according to the scientific question and the information it can reliably provide, rather than assumed to be routinely available for every candidate. Where reliable *operando* measurements cannot be obtained, standardized cyclic performance testing with calibrated gas analysis, complemented by pre- and post-cycling characterization, can still support candidate screening, although it cannot fully resolve transient structural or chemical changes during operation. Improving autonomous discovery thus requires not only higher measurement throughput, but also more reliable high-temperature experimental methodologies and robust interpretation of data acquired under realistic redox-cycling conditions.

**High-throughput redox testing**

Performance testing is a critical bottleneck that distinguishes redox-active oxide design from many conventional materials-discovery problems. Candidate materials must be evaluated across multiple metrics, including the oxygen storage capacity (**Fig. 2c**), redox kinetics, fuel yield, thermal-to-chemical energy conversion, cycling stability, and thermomechanical durability[13,34,184]. A staged testing strategy is therefore needed to prioritize measurements according to their relevance to material selection and the intended operating conditions. Initial screening should prioritize oxygen-exchange capacity and fuel yield under standardized redox conditions, together with the temperatures and atmospheres required to achieve these outputs. Among the available methods, thermogravimetric analysis (TGA), particularly when coupled with differential scanning calorimetry, gas chromatography or mass spectrometry, is well suited for early-stage screening because it requires only milligram-scale samples and quantifies mass change associated with oxygen release and uptake[185–187] (**Table 1**). With appropriate calibration, TGA data can be used to determine the oxygen nonstoichiometry, redox capacity, apparent reaction kinetics, and redox enthalpy[188,189]. Temperature-programmed reduction and oxidation further provide rapid proxy metrics for reducibility, oxygen release, re-oxidation behavior, and fuel-production activity under controlled gas environments[190,191]. Oxygen diffusion and surface exchange can be characterized by electrochemical impedance spectroscopy, electrical conductivity relaxation, and isotope-exchange techniques[192–195]. Thermomechanical stability can be evaluated separately using dilatometry, laser-flash analysis, thermal-shock testing, and attrition measurements[196–199].

Beyond initial performance, cyclic thermochemical materials must be evaluated for how their activity evolves during repeated redox operation. A high initial oxygen-exchange capacity may not translate into high cumulative fuel production if sintering, phase segregation, or structural degradation progressively reduces activity[200–202]. Consequently, candidates with similar initial oxygen capacities may require different follow-up tests if their degradation behavior or structural evolution differs during cycling. Accordingly, optimization targets should include cycle-dependent fuel yield, capacity retention, deactivation rate, and structural integrity over a defined operating window[203–205]. This challenge is not unique to thermochemical processes: heterogeneous catalysis and electrochemical energy conversion have also developed approaches for evaluating performance degradation and predicting durability over time. Methodologies developed in heterogeneous catalysis and electrochemical energy conversion, including accelerated stress testing[206,207], periodic diagnostic measurements[208], deactivation-kinetics analysis[209,210], and lifetime prediction[211,212], offer useful starting points. Their transfer to thermochemical systems nevertheless requires validation because accelerated protocols must preserve the relevant high-temperature redox and thermomechanical degradation mechanisms. Autonomous workflows should therefore combine rapid initial screening with selective extended-cycling tests, using early-cycle performance and uncertainty estimates to prioritize candidates for further testing, while reserving longer-term validation for promising candidates. Early-cycle results should guide test allocation rather than be treated as definitive evidence of long-term durability.

Fixed-bed and fluidized-bed tests provide reactor-scale validation under more representative heat- and mass-transfer conditions[213–215]. However, they remain substantially more complex and relatively immature for high-throughput testing because they require gram-scale samples and complex sample loading and removal[216,217]. This contrast motivates a tiered testing strategy: rapid TGA-based screening can be used to generate a shortlist, with reactor-based validation reserved for the most promising candidates. However, this tiered approach remains constrained: true high-throughput thermochemical redox testing is still limited by single-sample furnace architectures, complex gas switching, long cycling times, sample-handling constraints, and nonstandardized data processing. A transferable autonomous testing system would combine programmable atmosphere control, parallel or miniaturized reactors, online gas analysis, rapid mass- or heat-flow detection, and automated extraction of kinetic and degradation metrics. Such platforms would allow harsh operating conditions to become part of the discovery loop, rather than a late-stage validation step. Beyond hardware, integrating these measurements with standardized data processing and uncertainty-aware decision models would allow reactor-level evidence to inform subsequent material selection and testing, rather than being used solely for end-stage validation.

**Computation-guided screening of redox-active oxides**

High-throughput computation provides a complementary route to prioritize redox-active oxide candidates before experimental synthesis (**Fig. 2d**). For oxide materials, DFT, computational phase-diagram approaches (*e.g.*, calculation of phase diagrams), molecular dynamics, lattice dynamics, and thermodynamic modelling have been used to evaluate the phase stability[75,218], synthesizability[219,220], oxygen-vacancy formation energy[75,112,221], oxygen nonstoichiometry[112,222,223], redox free energy[57,73,74], oxygen diffusion[224–226], surface exchange[16,118,227], and thermal properties[228–230]. These descriptors are directly relevant to thermochemical fuel production because they link the composition and structure to the oxygen release, redox reversibility, phase evolution, and finite-temperature stability[61,75].

However, redox-active oxide prediction requires more than simple 0 K stability or isolated vacancy descriptors. Relevant performance depends on temperature, oxygen partial pressure, defect interactions, surface reactivity, metastability, phase transitions, and coupled chemical-mechanical evolution[40,61,231]. For example, convex-hull stability provides a useful first filter for synthesizability but does not capture the kinetic accessibility. The oxygen vacancy formation energy is a powerful descriptor of reducibility, but its value depends on the defect site, surface versus bulk environment, oxygen chemical potential, and finite-temperature corrections[75,219,232]. Likewise, the oxygen nonstoichiometry and redox free energy need to be evaluated under conditions that mimic actual reactor operation — specifically, within the relevant temperature and oxygen-partial-pressure windows. Future computational workflows should move beyond static descriptor screening toward condition-aware modeling under relevant thermochemical operating conditions[35,75]. Recent ML-DFT studies further suggest that screening should consider not only the reducibility, but also the thermal stability, mechanical robustness, and balance between the activity and stability[233,234].

**Machine learning, inverse design, and adaptive decision-making**

ML provides an integrative layer for connecting computation, synthesis, characterization, and testing within autonomous discovery workflows[235]. In thermochemical fuel production and related oxide systems, ML models have been used to predict the reactivity of redox-active oxides[236,237], oxygen storage capacity[238], vacancy formation energies[239], oxygen nonstoichiometry[45,223], diffusion behavior[240], and process-level performance[241]. Inverse-design strategies, including genetic algorithms and surrogate-model-guided searches, further enable direct optimization toward targeted redox properties or oxygen-release behaviors[242]. ML interatomic potentials extend these capabilities by enabling near-DFT-accuracy simulations of vacancy formation, diffusion, and structural evolution at larger length and time scales[243]. Recent chemical-looping-related models span redox-active oxide design, argon purification, biomass chemical looping, and balanced reactivity-stability screening. However, most studies addressed individual prediction tasks; they have not yet demonstrated a unified autonomous workflow[233,244].

For autonomous redox-active oxide discovery, the most important role of ML is not prediction alone but adaptive decision-making[127,245]. Models should quantify uncertainty, identify the most informative next experiments, learn from failed candidates, and balance competing objectives such as the reactivity, oxygen capacity, transport kinetics, phase stability, and thermomechanical durability[45,246]. This capability is particularly important because the best redox-active oxides are unlikely to maximize any single descriptor. Instead, they should occupy Pareto-optimal regions in which the activity and durability are jointly optimized under realistic cycling conditions[57,93]. Active-learning policies should also incorporate failed syntheses, unmeasurable candidates, and safety-constrained experiments, as these outcomes help define the feasible region of the redox-active oxide design space[247–249].

The maturity of multi-objective optimization differs considerably between thermochemical fuel production and adjacent materials fields. For thermochemical materials, computational screening has combined descriptors such as oxygen-vacancy energetics, phase stability, and electronic properties to identify promising redox-active oxides, followed by experimental validation of selected candidates[75,80,108]. More recently, ML-assisted screening and human-guided compositional refinement have further demonstrated how computational predictions and experimental feedback can improve candidate selection and material stability1. However, these studies have primarily involved multi-criteria screening or human-guided optimization rather than autonomous experimental feedback loops that jointly optimize multiple performance objectives. In contrast, self-driving laboratories in adjacent materials fields have demonstrated Pareto-front exploration of conductivity and processing temperature in palladium films[250] and constrained multi-objective Bayesian optimization of silver nanoparticle synthesis[251]. Translating these capabilities to thermochemical fuel production remains challenging because high-temperature testing is relatively slow, durability requires extended cycling, and performance objectives such as oxygen capacity, fuel yield, reaction kinetics, and structural stability often require different measurements. Integrating these measurements into reliable, uncertainty-aware experimental feedback loops represents an important next step toward autonomous multi-objective optimization of redox-active oxides.

A practical autonomous workflow should therefore prioritize measurements and computational analyses according to their decision value, rather than applying every available technique to every candidate. As detailed in the preceding subsections, relatively accessible computational filters for phase stability and synthesizability can narrow broad chemical spaces, followed by rapid, standardized experimental screening of oxygen exchange and fuel production. Only candidates that pass these initial criteria need to be advanced to more targeted characterization, mechanistic analysis, and extended cycling, with resource-intensive reactor-level validation reserved for materials whose performance, cycling behavior, and synthesis reproducibility justify the additional effort. Progression between these stages should not rely solely on fixed thresholds: uncertainty estimates and the expected value of additional information can help determine whether a candidate should advance, undergo further characterization, or be deprioritized. In this framework, AI and automation are most useful for coordinating workflows, processing data, quantifying uncertainty, prioritizing experiments, and updating models from experimental feedback. They cannot compensate for unreliable measurements, inadequate high-temperature reactors, poor gas control, temperature nonuniformity, or insufficiently validated accelerated tests; these limitations require advances in instrumentation, reactor engineering, calibration, and experimental protocols. Thus, the most effective route toward autonomous optimization is not maximal automation or exhaustive measurement, but a staged allocation of computational and experimental resources in which increasingly demanding methods are used only when they can materially improve decisions about candidate selection and validation.

## CHALLENGES AND PROSPECTS FOR AUTONOMOUS REDOX-ACTIVE COMPLEX OXIDE DISCOVERY

Accelerating redox-active oxide discovery requires more than simply transferring high-throughput tools to thermochemical fuel production. The challenge spans several levels (**Fig. 3**). Experiments would benefit from miniaturization, parallelization, and automation at high temperatures and in the presence of reactive gases. Computations need to better account for finite-temperature redox thermodynamics, defect chemistry, and kinetic accessibility. ML models should handle sparse, noisy, and history-dependent datasets. Data infrastructures should connect synthesis, characterization, testing, and modeling in interoperable, machine-readable formats. Progress in these areas is needed to move autonomous discoveries beyond isolated demonstrations toward integrated, closed-loop platforms for thermochemical fuel production.

### Experimental automation under extreme redox conditions

Automating experiments for thermochemical fuel production is challenging because redox-active oxides span broad chemical and structural spaces while operating under demanding redox environments.

For synthesis, throughput remains a major challenge. The large design space of thermochemical fuel production materials calls for synthesis strategies that can rapidly generate candidate libraries, rather than exhaustively optimizing each composition at the earliest stage. These platforms should also remain cost-effective because redox-active oxides are ultimately needed in large quantities. At the same time, synthesis workflows need sufficient chemical flexibility to accommodate diverse oxide families, including single-metal oxides, spinels, perovskites, mixed ionic–electronic conductors, and selected polyanionic compounds. Morphological control is also important because the particle size, porosity, surface area, and microstructure influence the oxygen-transport kinetics, sintering resistance, and mechanical durability. Automated synthesis routes should therefore be reproducible, tolerant to process variability, and compatible with robotic handling, standardized substrates, and unattended operation. Chip-based sample arrays can reduce sample consumption, provide consistent preparation histories, and support high-throughput synthesis.

For characterization, the key difficulty is obtaining meaningful descriptors under conditions relevant to thermochemical redox operation. Automated workflows should accommodate high temperatures and controlled gas atmospheres, which adds complexity to the sample environments, gas handling, calibration, and instrument control. Rapid measurements should still provide interpretable descriptors, such as the oxygen nonstoichiometry, phase evolution, defect populations, valence changes, and microstructural degradation. Diffraction, spectroscopy, and microscopy also generate high-dimensional datasets that require automated and uncertainty-aware analysis. Because the state of a redox-active oxide depends strongly on the synthesis route and redox history, standardized protocols and metadata are needed for meaningful comparison across samples. More robust algorithms are still needed for automated phase identification, spectral interpretation, and defect quantification under *in-situ* or *operando* conditions. *Operando* neutron diffraction and complementary local-structure probes can track the oxygen exchange, phase evolution, and reactor state under working redox conditions[175–177].

For testing, the main challenge is reproducing thermochemical environments in a format compatible with automation and high throughput. Two-step thermochemical fuel production requires high temperatures and cyclic switching between reducing and oxidizing conditions, which are generally implemented and monitored *via* TGA to evaluate the oxygen exchange capacity. Although TGA offers a high degree of automated testing, its intrinsically low testing throughput limits comprehensive evaluation of the long-term cyclic stability and sintering resistance. Reactor-level redox cycling tests provide more practical evaluations, especially regarding attrition and thermochemical degradation. However, such reactor-level platforms typically suffer from low levels of automation, high labor requirements, and low throughputs. Long-term redox cycling also places substantial demands on reactor materials, seals, sensors, and control hardware[31,252]. Miniaturized reactors and sensors can improve the throughput, but they should preserve the thermal uniformity, gas–solid contact, and measurement fidelity. Parallel testing can further introduce thermal cross-talk, gas leakage, and cross-contamination, especially at elevated temperatures. Because thermochemical fuel production is cyclic rather than purely steady-state, automated platforms require coordinated control of the temperature, gas switching, and data acquisition. Capturing long-term degradation, including sintering, attrition, and phase separation, within accelerated yet realistic protocols remains difficult. A practical approach is to use standardized short-cycle protocols for initial screening, followed by accelerated stress tests and extended cycling for shortlisted candidates. These accelerated protocols should be benchmarked against conventional cycling to verify that they preserve the degradation mechanisms relevant to realistic operating conditions. Autonomous testing platforms should therefore report not only the sample throughput but also the uncertainty, calibration history, and operational constraints that determine whether parallel measurements are comparable[69,253,254].

These bottlenecks require complementary rather than interchangeable solutions. AI and automation can coordinate experimental sequences, standardize data processing, extract performance and degradation metrics, quantify uncertainty, and prioritize subsequent experiments. They cannot, however, compensate for inadequate reactor design, unreliable seals or sensors, poor temperature uniformity, gas leakage, or insufficient measurement fidelity. Addressing these constraints requires improvements in reactor and instrument design, calibration, and experimental protocols. Similarly, miniaturized and parallelized testing configurations, as well as accelerated screening protocols, must be experimentally validated to ensure that gains in throughput do not compromise measurement comparability or alter the degradation mechanisms relevant to realistic thermochemical operation.

The practical value of an autonomous discovery platform should be evaluated at the platform level rather than inferred solely from the throughput of individual experiments. Initial investment in instrumentation, high-temperature reactors, gas-control systems, robotic handling, software integration, and safety infrastructure can be substantial, while workflow development and commissioning may require additional time and resources. The benefits are therefore most likely to justify these investments when large or recurring screening campaigns use sufficiently standardized workflows, allowing fixed costs to be distributed across many experiments and projects. Higher platform utilization, reduced manual intervention, improved experimental reproducibility, and faster identification of promising candidates may further improve the overall value of automation. Conversely, for small or infrequent campaigns, fully integrated platforms may be less cost-effective, making partial automation or shared facilities more practical alternatives. Thus, the appropriate level of automation should be determined by the expected screening demand, workflow complexity, and long-term operational benefits, rather than by throughput alone.

**Computation beyond static descriptor screening**

Computational screening can help narrow the large chemical space of redox-active oxides, but several factors still limit its predictive power. Synthesizability cannot be inferred from thermodynamic stability alone. Convex-hull distances and phase-diagram calculations are useful filters, but they do not capture kinetic accessibility, metastable intermediates, nucleation barriers, interfacial energies, or solid-state diffusion pathways. These factors can strongly affect whether a predicted oxide can be synthesized and retained during redox cycling. Thermochemical fuel production operates at high temperatures, where 0 K total energies are often insufficient. More reliable predictions require Gibbs free energies that include the relevant gas-phase, configurational, vibrational, and electronic entropy contributions. This becomes especially important for nonstoichiometric oxides, where oxygen-vacancy populations, mixed-valence states, and phase stability vary with the temperature and oxygen partial pressure. Predicting the oxygen nonstoichiometry therefore requires models that account for defect–defect interactions, local coordination environments, and finite-temperature thermodynamics beyond the dilute-vacancy approximation. The oxide redox performance is inherently multiscale. Redox activity depends on the thermodynamics, defect chemistry, surface reactions, bulk oxygen diffusion, microstructural evolution, and mechanical degradation. Capturing all these processes with first-principles calculations alone remains computationally expensive. A more practical path is hierarchical modelling that combines high-throughput descriptors, thermodynamic models, kinetic simulations, ML interatomic potentials, and uncertainty-aware surrogate models. Such workflows could prioritize candidates based on the predicted activity, synthesizability, stability, and robustness within realistic operating windows. Foundation models and ML interatomic potentials may accelerate structure generation and atomistic screening, but their predictions should still be evaluated against finite-temperature thermodynamics, gas-atmosphere constraints, and cyclic *operando* validation before candidates are considered viable redox-active oxides[93,255–257].

Condition-aware computational modelling evaluates oxide thermodynamics, defect chemistry, and redox behavior under operating conditions relevant to thermochemical fuel production, rather than relying solely on static descriptors calculated at 0 K. Relevant conditions include reduction and oxidation temperatures, oxygen partial pressures, gas compositions (such as $H_2O/H_2$ and $CO_2/CO$ ratios), and the duration of individual redox steps. Reactor-level factors, including gas flow, residence time, and temperature gradients, further influence the local conditions experienced by the material. Several computational capabilities are already available for evaluating these effects. Beyond conventional DFT-based screening of oxygen-vacancy formation energies and phase stability, finite-temperature thermodynamic and defect-chemical models have been developed to predict oxygen non-stoichiometry, phase equilibria, and redox thermodynamics as functions of temperature and oxygen partial pressure[51,258,259] . These approaches provide a basis for screening candidates within specified thermochemical operating windows.

However, thermodynamic predictions alone do not fully describe material performance during finite-duration redox cycling. Capturing transient behavior requires coupling equilibrium thermodynamics with surface-exchange kinetics, bulk oxygen diffusion, and defect interactions, while longer-term performance predictions must also account for phase evolution, microstructural changes, and degradation[260–262]. Integrating these processes across different length and time scales remains computationally challenging, particularly for compositionally complex oxides. A practical route forward is therefore to combine computational models of different fidelities, using relatively inexpensive descriptors for initial screening and more detailed condition-aware and kinetic models for selected candidates. Their integration with experimentally validated cycling data and uncertainty-aware decision-making could progressively support autonomous optimization under realistic thermochemical operating conditions.

**Machine learning and artificial intelligence for adaptive decision-making**

ML can help transform high-throughput workflows into more adaptive discovery systems, but its application to thermochemical fuel production remains challenging. Experimental datasets are often sparse, heterogeneous, and history-dependent, with the measured performance shaped by the composition, synthesis route, microstructure, temperature, gas atmosphere, and cycling protocol. This heterogeneity reflects the diversity of the field: water splitting, $CO_2$ splitting, and syngas production have been studied using different reactor configurations, operating atmospheres, temperature windows, sample morphologies, cycling protocols, and performance metrics[13,57]. As a result, literature data are difficult to integrate into unified databases, and cross-study comparability remains limited, which constrains the transferability of ML models across materials and operating conditions. Active-learning algorithms should therefore balance exploration of broad chemical spaces with the exploitation of promising regions, while accounting for experimental noise, drift, irreversible degradation, and safety constraints under high-temperature operation. Progress would benefit from more standardized testing protocols, including reference conditions, benchmark reactions, minimum reporting requirements, and common performance indicators. A practical data strategy should also distinguish literature-mined data, first-principles data, automated synthesis records, *operando* characterization streams, reactor-testing data, and process simulations, while preserving failed syntheses, infeasible candidates, and unmeasurable conditions as informative outcomes[263–265].

Another important challenge is connecting prediction to synthesis. A model may identify a promising target composition, but autonomous optimization also requires a feasible route to prepare it with the desired phase purity, oxygen stoichiometry, and morphology. This requirement underscores the need for inverse-design approaches that couple composition optimization with retrosynthesis planning. This coupling creates a need for materials retrosynthesis and process-inference models that recommend not only compositions but also precursors, thermal profiles, atmospheres, and processing conditions. For redox-active oxides, such models should account for kinetic and thermodynamic constraints, which are often underrepresented in existing synthesis datasets[70,254,266]. Automated interpretation of experimental data is equally important. *In-situ* diffraction, spectroscopy, microscopy, and thermochemical measurements generate high-dimensional, time-resolved datasets with overlapping signals and evolving baselines. AI models should extract physically meaningful descriptors, such as phase fractions, valence changes, oxygen-release profiles, degradation rates, and kinetic parameters, while quantifying the uncertainty. These models approximate the mapping from raw measurement signals to descriptors that would otherwise require manual analysis or more computationally intensive interpretation. The extracted descriptors can then serve as inputs to performance-prediction models that relate material composition, synthesis history, experimentally derived descriptors, and operating conditions to target outcomes, including oxygen-release capacity, fuel yield, reaction kinetics, and cycling stability. In this second task, surrogate models approximate the input-output relationships learned from prior experimental measurements or computational evaluations, rather than mapping raw measurement signals directly to performance indicators. Once adequately trained and validated, these models can estimate outcomes for new candidates more rapidly than repeating the corresponding full experimental evaluations or computational calculations. The distinction between signal interpretation and performance prediction is important because errors in extracted descriptors can propagate into subsequent performance estimates and, ultimately, lead to misleading optimization decisions. Physics-informed models, interpretable descriptors, and thermodynamic consistency constraints can help maintain physically plausible predictions, while confidence intervals or posterior distributions can help prevent downstream optimization from being guided by overconfident fits. For reactor-level applications, performance predictions should be validated under representative operating conditions, because materials-level descriptors and performance estimates alone may not fully capture heat and mass transfer, reactor configuration, and other system-level effects.

**Data and system integration**

Another major challenge is integration. Many relevant tools already exist in partial form, but they remain fragmented across synthesis, characterization, testing, computation, and modeling. Autonomous discovery requires these modules to operate as connected workflows in which samples, metadata, models, and decisions remain synchronized. This synchronization is particularly difficult for thermochemical fuel production because the field spans materials science, catalysis, thermochemistry, reactor engineering, and computational modeling, each with different data conventions and reporting practices. Data standardization is therefore not simply an infrastructure detail, but a prerequisite for reliable closed-loop discovery. Useful datasets should include machine-readable records of compositions, precursor chemistries, synthesis routes, processing histories, sample geometries or morphologies, reactor configurations, gas atmospheres, temperature programs, cycling protocols, characterization conditions, data-processing workflows, and uncertainty. Establishing shared schemas, ontologies, and reporting standards for redox-active oxide research would help build transferable, ML-ready datasets. At a minimum, such datasets should report the composition normalization, precursor identity and purity, processing atmosphere, temperature calibration, gas composition and oxygen partial pressure, switching sequence, sample mass and geometry, cycle number, instrument model, and uncertainty estimates[253,264,267].

Beyond individual platforms, progress would benefit from broader adoption of FAIR data principles, which make data findable, accessible, interoperable, and reusable[268,269]. Standardized metadata, common reporting schemes, and open repositories tailored to thermochemical fuel production could allow data generated by different laboratories and platforms to be compared and reused more effectively. Such an infrastructure would turn individual experiments into cumulative knowledge assets, improving reproducibility and enabling autonomous systems to learn from shared community data rather than isolated datasets alone. At the platform level, future systems may include AI agents that coordinate experimental planning, synthesis, testing, characterization, and computation. These agents would need to reason across incomplete data, select experiments under uncertainty, adapt to hardware or safety constraints, and update models as new results emerge[270]. In practice, such systems are best viewed as orchestration layers coupled to constrained optimizers and human-in-the-loop safety checks, rather than fully autonomous black boxes[69,253].

The key challenge for autonomous redox-active oxide discovery is the translation and integration of existing tools into platforms designed for the thermochemical realities of fuel production. Future workflows should prepare chemically diverse oxides, test them under controlled high-temperature redox environments, characterize their evolving structures and chemistries, and use ML to guide subsequent experiments. By combining extreme-environment automation, condition-aware computation, physics-informed AI, and standardized data infrastructure, redox-active oxide development can move toward adaptive, data-rich, and mechanism-guided optimization. This transition will be important for developing materials that are active, durable, scalable, and deployable under realistic thermochemical conditions.

## CONCLUSIONS

Thermochemical fuel production is critical to global decarbonization, yet its advancement is constrained by severe materials challenges. Its progress depends, in part, on redox-active oxides capable of sustaining oxygen exchange under high temperatures, reactive atmospheres, and repeated redox cycling, while simultaneously balancing the thermodynamics, kinetics, defect chemistry, microstructure, and thermomechanical robustness. These coupled requirements make the design of redox-active oxides difficult to address through conventional trial-and-error experimentation or static, single-objective screening alone. Autonomous and data-driven materials strategies offer a promising route to accelerate the design and optimization of complex redox-active oxides. By integrating high-throughput computation, automated synthesis and testing, advanced characterization, ML, and AI-guided decision-making into closed-loop workflows, such approaches can help explore broad chemical spaces, evaluate candidates under relevant operating conditions, learn from both successful and unsuccessful outcomes, and progressively refine design rules. Their value lies not in replacing physical understanding but in making materials optimization more systematic, data-rich, and adaptive. Looking forward, continued progress will require the convergence of extreme-environment automation, condition-aware computation, physics-informed AI, and a standardized data infrastructure. Together, these capabilities could shift redox-active oxide development from empirical optimization toward reproducible, mechanism-guided, and self-improving workflows. Realizing this transition, alongside continued advances in reactor and process design, will be essential for translating two-step thermochemical fuel production into scalable technologies with meaningful climate and industrial impacts.

## AUTHOR CONTRIBUTIONS

Author Contributions: Z.Y.: Conceptualization, Methodology, Supervision, Project Administration, Funding Acquisition, Writing – Review & Editing. S.G.: Investigation, Data Curation, Visualization, Writing – Original Draft. M.L.: Formal Analysis, Writing – Review & Editing. H.H.: Writing – Review & Editing. Z.Z.: Methodology, Investigation, Software, Writing – Original Draft. Y.L.: Methodology, Writing – Original Draft. X.L.: Data Curation, Formal Analysis. C.F.: Methodology, Investigation. J.D.: Data Curation. J.H.: Supervision, Writing – Review & Editing. W.G.: Supervision, Writing – Review & Editing. Y.Y.: Supervision, Writing – Review & Editing. T.D.: Methodology, Supervision, Writing – Review & Editing. C.D.: Supervision, Writing – Review & Editing. R.C.: Conceptualization, Methodology, Supervision, Writing – Review & Editing. All authors contributed to the preparation of this manuscript.

## CONFLICTS OF INTEREST

The authors declare no conflict of interest.

## ACKNOWLEDGEMENTS

Z.Y., S.G., Z.Z., Y.L., X.L., C.F., and J.D. were supported by the National Natural Science Foundation of China (Grant NO. 52373228). M.L. and R.C. were supported by the National Natural Science Foundation of China (Grant NO. 52506156).

## DATA AVAILABILITY

The data supporting the analysis and discussion presented in this Perspective were compiled from the published literature cited herein. No new experimental data, software, or code were generated for this work.

## Figures

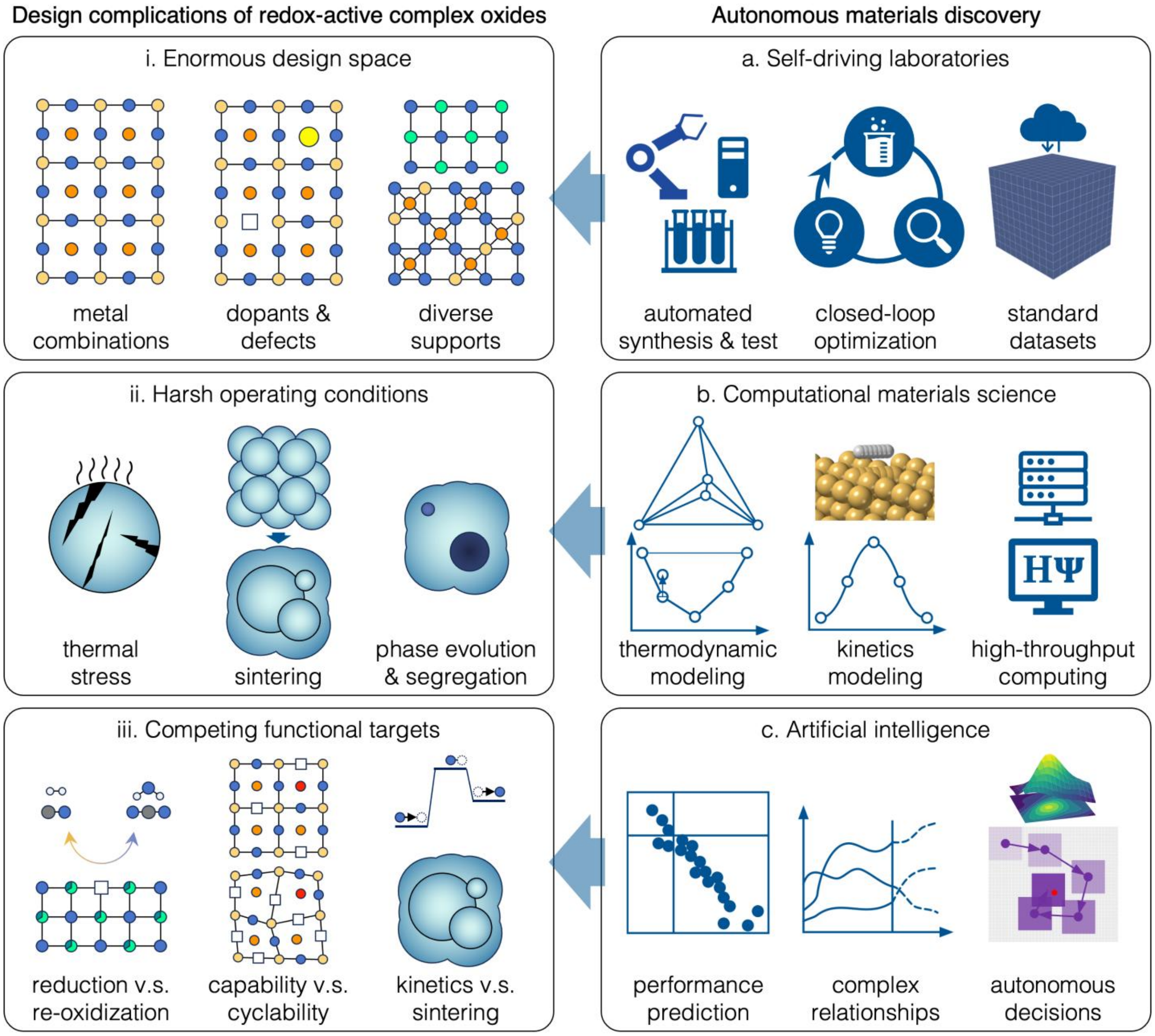

**Figure 1. Data-driven materials discovery for addressing the design challenges of redox-active materials for thermochemical fuel production.**

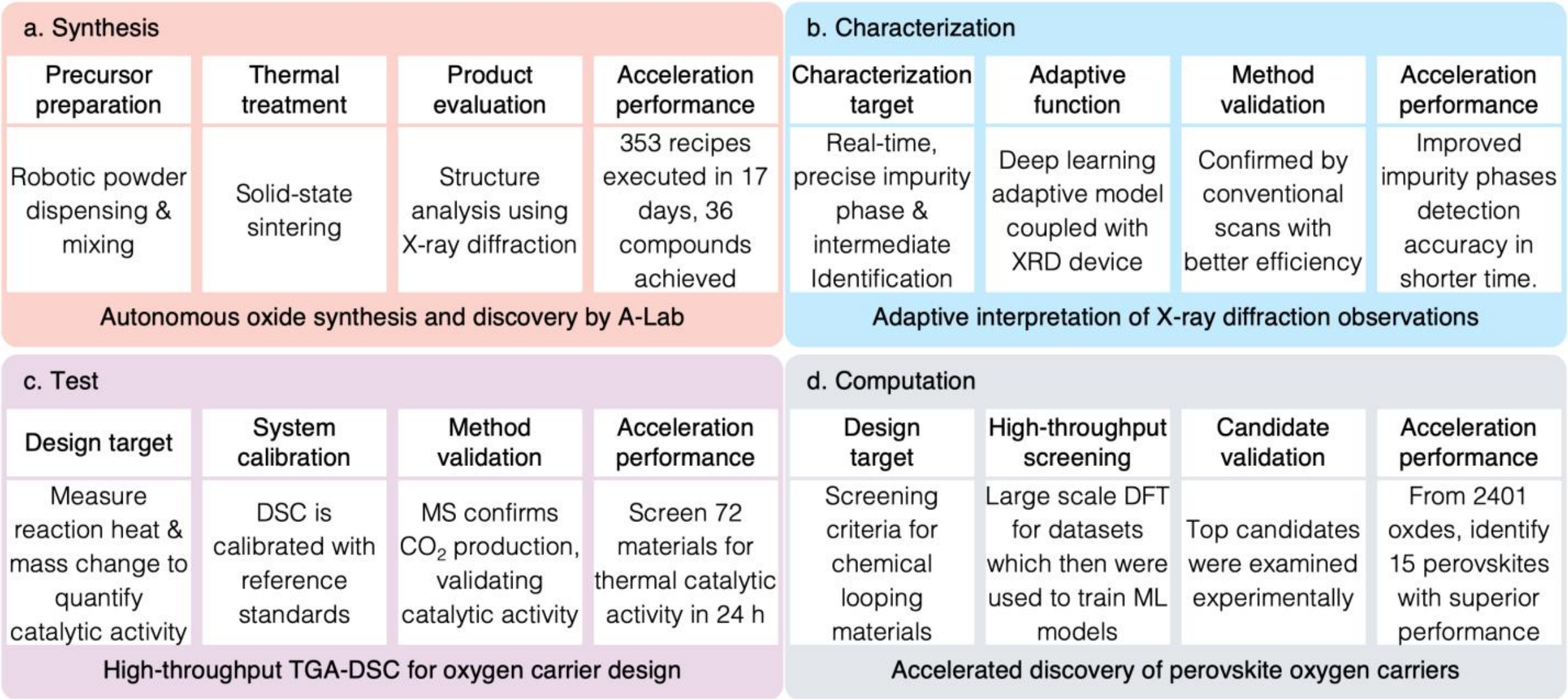

**Figure 2. Examples of state-of-the-art data-driven materials discovery.** a. Autonomous oxide synthesis and discovery by autonomous lab[70]. b. Adaptive interpretation of X-ray diffraction observations[117]. c. High-throughput TGA-DSC for redox-active oxide design[271]. d. Accelerated discovery of perovskite oxides[57].

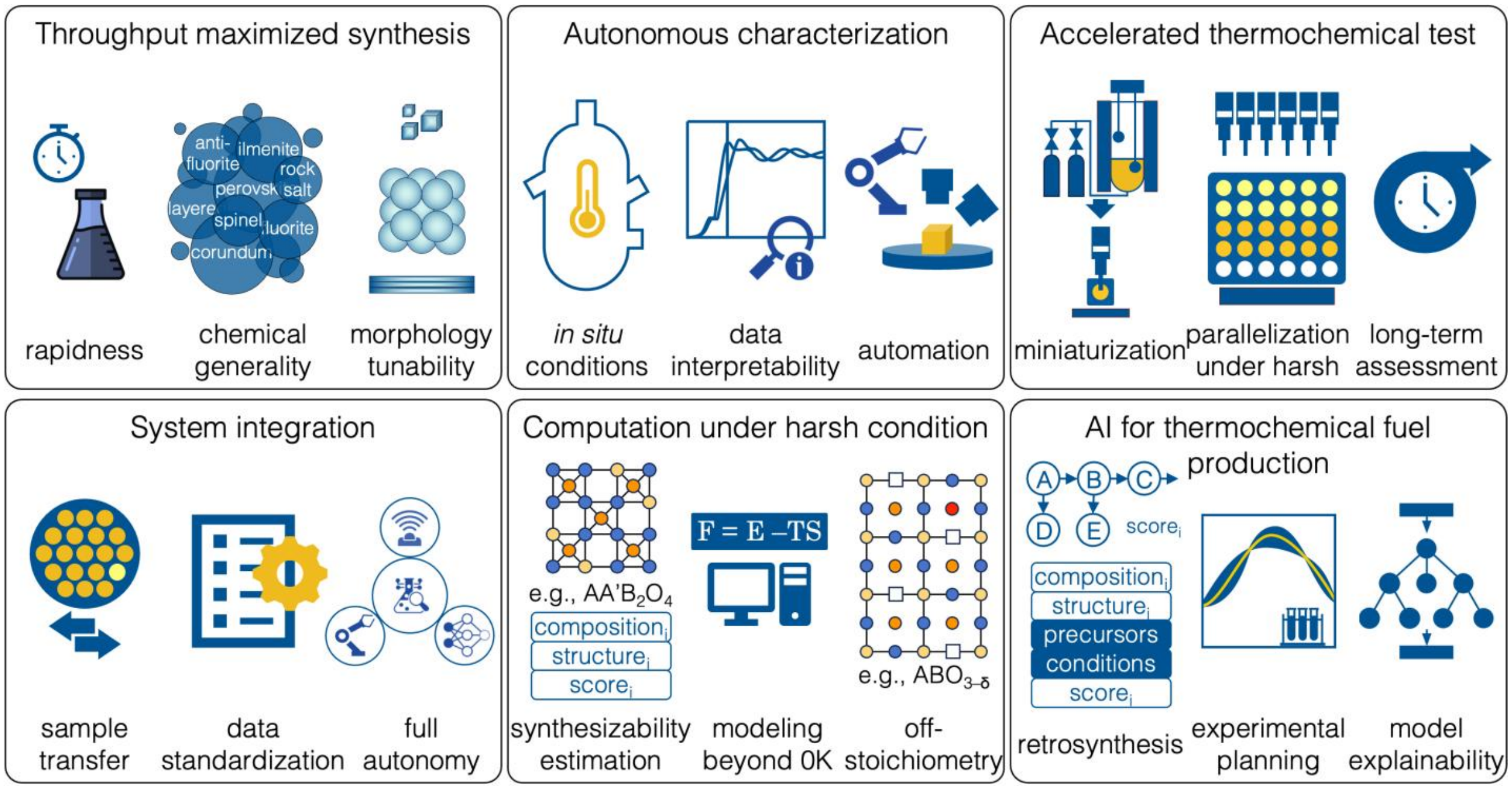

**Figure 3. Capability roadmap for autonomous materials discovery and optimization in thermochemical fuel production.**

## Tables

Table 1. Accelerated experimental and computational capabilities for autonomous optimization of complex oxides

| Capability | Author | Materials system | Approach | Performance | Autonomy level |
|---|---|---|---|---|---|
| Synthesis | Wang et al.[129] | Oxide ceramics | Flash Joule heating for ultrafast high-temperature sintering | Enabled synthesis and sintering of bulk oxides on a timescale of seconds | L1 |
| | Hampson et al.[130] | Oxides | Automated high-throughput solid-state synthesis workflow with XRD characterization | Synthesized 192 oxide samples within 3 weeks | L1 |
| | Omidvar et al.[131] | Perovskite solid solutions | ML-guided screening with automated synthesis and characterization | Enabled rapid experimental validation of ML-selected perovskite candidates | L2 |
| | Pelkie et al.[132] | Nanoporous silica | Automated sol–gel synthesis integrated with SAXS | Synthesized and characterized 63 samples within 24 h | L1 |
| | Engel et al.[133] | Oxide nanoparticles | Automated robotic flame spray pyrolysis (AutoFSP) | Reduced operator workload by 2–3× and achieved composition control error within ±5% | L1 |
| | Weidenhof et al.[136] | Oxide nanoparticles | High-throughput liquid-feed FSP synthesis and catalytic screening | Rapid screened ~40 oxide nanoparticle catalysts | L1 |
| | Heo et al.[145] | Ce-substituted $(Ba,Sr)MnO_3$ perovskites | Combinatorial pulsed laser deposition synthesis with spatially resolved XRD/XRF characterization | Ce-substituted $(Ba,Sr)MnO_3$ candidates exhibited ~2–3 × higher $H_2$ production than $CeO_2$ | L1 |
| Characterization | Park et al.[147] | Inorganic materials | Automated high-throughput XRD/XRF platform (MAXIMA) | Acquired and processed 162 spatially resolved XRD/XRF datasets in ~94 min | L1 |
| | Oji et al.[272] | Inorganic materials | Automated XAFS measurements with robotic sample handling and alignment | Enabled automated XAFS measurements for up to 80 samples | L1 |
| | Olszta et | Inorganic | Automated STEM | Enabled automated | L1 |

| Capability | Author | Materials system | Approach | Performance | Autonomy level |
|---|---|---|---|---|---|
| | al.[110] | materials ($MoO_3$ flakes) | acquisition with few-shot ML-based feature classification | feature mapping and statistical characterization; providing a basis for feature-directed adaptive sampling | |
| | Kirkham et al.[273] | Oxygen-deficient perovskites | Automated gas environment system coupled with *in-situ* neutron diffraction | Enabled automated *operando* structural measurements under controlled reactive environments | L1 |
| | Hao et al.[111] | Single-crystal materials | ML-assisted automated single-crystal neutron diffraction | Enabled automated experiment planning, peak recognition, and data analysis | L1 |
| Testing | Quayle et al.[137] | Rare-earth-doped ceria–zirconia oxides | High-throughput robotic synthesis and automated oxygen storage characterization | Developed a proxy model for rapid prediction of oxygen storage capacity | L2 |
| | Li et al.[167] | $H_2O_2$-containing solutions | High-throughput robotic colorimetric titration with computer vision | Enabled automated endpoint detection using computer vision | L1 |
| | Loskyll et al.[271] | Oxide-based heterogeneous catalysts | High-throughput catalytic screening using simultaneous TGA and DSC | Screened ~70 catalysts per day | L1 |
| | Liu et al.[274] | $CaMn_{0.5}Ti_{0.375}Fe_{0.125}O_3$ perovskite oxygen carrier | Micro-fluidized-bed thermogravimetric analysis (MFB-TGA) based on real-time mass measurement of fluidized particles | Enabled high-throughput kinetic analysis; $O_2$ release and uptake rates ~2× and ~4× higher, respectively, than those measured by conventional TGA | L1 |
| | Pishahang et al.[275] | $CaTi_xMn_{0.9-x}Mg_{0.1}O_3$ (x = 0, 0.125) perovskite oxygen carriers | SETSYS evolution TGA coupled to a fully automated in-house gas-mixing system | Fully automated gas switching enabled programmed redox cycling; 39 cycles were performed at each temperature from 800 to 1000 ℃ (195 cycles in total). | L1 |

| Capability | Author | Materials system | Approach | Performance | Autonomy level |
|---|---|---|---|---|---|
| | Dubey et al.[276] | Natural calcined and synthetic ilmenite oxygen carriers | Batch fluidized-bed reactor with computer-programmed automated cycling; online gas analysis and data logging | Computer-controlled atmosphere switching enabled automated redox cycling; 33 redox cycles were conducted for alkali-fuel experiments. | L1 |
| Computation | Emery et al.[75] | $ABO_3$ perovskites | High-throughput DFT screening using $E_{V,O}$ and stability descriptors | Screened 5329 perovskites using >11000 DFT calculations and identified 139 candidates for solar thermochemical hydrogen (STCH) | L2 |
| | Zhang et al.[34] | Compositionally complex perovskite oxides | DFT-guided Monte Carlo optimization with experimental validation | Identified a perovskite composition with 89.97 $mmol/mol_{oxide}$ $H_2$ production and >50 redox cycles | L2 |
| | Baldassarri et al.[221] | $ABO_3$ perovskite oxides | High-throughput DFT+U screening of oxygen vacancy thermodynamics | Predicted $\Delta E_{V,O}$ with 0.2–0.6 eV/O agreement with experimental reduction enthalpies | L2 |
| | Morelock et al.[80] | Mixed Mn/Ni perovskites | DFT-guided screening and experimental validation of double perovskites | Identified candidates achieving ~65 µmol/g $H_2$ per cycle under concentrated steam conditions | L2 |
| | Park et al.[114] | Gd-containing perovskite oxides | DFT-guided screening with experimental validation | Identified $Gd_{0.5}La_{0.5}Co_{0.5}Fe_{0.5}O_3$ with $H_2$ yields of 101–141 µmol/g | L2 |
| Machine learning | Witman et al.[112] | Structurally diverse oxides | Defect graph neural network (dGNN) prediction of oxygen vacancy formation enthalpy | Predicted oxygen vacancy formation enthalpies across diverse oxides and enabled STCH materials screening | L2 |
| | Clauser et al.[1] | $BaFe_2O_4$-based oxides | dGNN-guided screening with experimental validation and human-in-the-loop optimization | Identified $BaFe_2O_4$ as an STCH candidate and improved its thermal stability through Al | L2 |

| Capability | Author | Materials system | Approach | Performance | Autonomy level |
| --- | --- | --- | --- | --- | --- |
| | | | | substitution | |
| | Perry et al.[116] | Perovskite oxides | Random-forest-guided screening with DFT and experimental validation | Screened 6264 compositions and identified a candidate with ~250 K lower reduction temperature than that of $CeO_2$ | L2 |
| | Bare et al.[115] | Gd-containing perovskites | ML- and DFT-based high-throughput screening with experimental validation | Screened 4392 compositions, identified 83 candidates, and experimentally validated three STCH-active oxides | L2 |
| | Douglas et al.[49] | Structurally diverse metal oxides | dGNN-guided oxygen vacancy prediction with DFT screening and experimental validation | Identified 12 candidates; $Sr_3PrMn_2O_8$ achieved ~30% higher $H_2$ yield than $CeO_2$ at 1673 K | L2 |